\documentclass{elsartmod}

\usepackage{epsfig}

\usepackage{amssymb}

\newtheorem{theorem}{Theorem}[section]
\newtheorem{lemma}{Lemma}[section]
\newtheorem{remark}{Remark}[section]

\begin{document}

\begin{frontmatter}

\title{On controllability and observability of chains formed by point masses connected with springs and dashpots}
\author{Adriano MONTANARO}

\address{Dep. Metodi e Modelli Matematici per le Scienze Applicate\\
University of Padua \\via Trieste, 63,  35121, Padova, Italy
\\ Email: montanaro@dmsa.unipd.it}

\begin{abstract}
We consider a physical system constituted by a finite chain of point masses consecutively linked by linear springs and dashpots.  At one of the end points acts an external control force aligned with the chain and the system is observable by the position of the other end point.
We show that, whatever is the number of the point masses, if  the sequence of the elastic constants is proportional to the sequence of the dashpot constants, then the mechanical system is completely controllable,  completely observable,  completely reachable and completely re-construictible, in the sense of control theory.
\end{abstract}

\end{frontmatter}

\section{Introduction}
\subsection{Premise}                                                                  \label{subsection:premise}
In  \cite{M:ControlSprings}
a finite chain of mass points consecutively linked by linear springs is considered; the mechanical system  is externally controlled by a control force aligned with the chain, applied at one of the end points, and is observable by the position of the other end point. 
There it is shown that such a system is completely controllable,  completely observable,  completely reachable and completely re-construictible, in the sense of control theory, for all choices of the physical parameters.

Here we consider a spring-mass system similar to the one described above but includes viscous damping which is caused by the presence of dashpots connecting consecutive points of the chain. 
We show that such a system is completely controllable,  completely observable,  completely reachable and completely re-construictible provided the sequence of the dashpot constants is proportional to the sequence of the elastic constants. This is true for any choice of the number of points, of their masses, and of the elastic constants of the springs.

\subsection{On linear systems with control inputs}                                                    \label{subsection:lin3}

Following the same notations of \cite{FM:Appunti}, we consider the continuous linear first order system 
%%%%%%%%%%%%%%%%%%%%%%%%%%%%%%%%%%%
\begin{eqnarray}  																										
\dot{\bf z} ={\bf F}\,{\bf z}+{\bf g}\,{\bf u} \; ,              \label{eqnarray:sy1}   \\ 
{\bf y} ={\bf h}\,{\bf z}+{\bf d}\,{\bf u} \; ,                  \label{eqnarray:sy2}
\end{eqnarray}
%%%%%%%%%%%%%%%%%%%%%%%%%%%%%%%%%%%%%
where ${\bf z}$,  ${\bf u}$ and ${\bf y}$ are real vectors of respective dimensions $n$, $m$ and $p$, 
$\dot{\bf z}={{d{\bf z}}/{dt}}$,  
${\bf F}$ is the $(n \times n)$ state matrix,  ${\bf g}$ is the $(n \times m)$ input matrix, 
${\bf h}$ is the $(p \times n)$ output matrix,  and ${\bf d}$ is the $(p \times m)$ direct transmission matrix.
In literature the two equations above are usually referred as {\it state equation} and {\it output equation}, respectively.

% CONTROLLABILITY:  A state is said to be controllable if there is some smooth enough control u(t) which is capable of changing the state from any given initial value to any desired final value in a finite lenght of time.   If alla states are controllable then the system is said to be controllable.

%Similarly, a system is said to be observable if every initial state (x(0) can be determined by observing the system output  y over some finite time period.

A mechanical system $\, \Sigma\,$ represented by equations (\ref{eqnarray:sy1})-(\ref{eqnarray:sy2}) is said to be
(\cite{C:LinearDs}--\cite{R:TeoriaDeiSistemi})

$(A)\;$ {\it completely reachable}                        
if for each state  ${\bf z}$ 
there are an instant $t>0$ and an input function $u(.)$, defined in $[0, \, t]$,  such that the solution ${\bf z}(.)$ to equation  (\ref{eqnarray:sy1}) satisfies
${\bf z}(0)={\bf 0}$, $\;{\bf z}(t)={\bf z}$;
 
$(B)\;$  {\it completely controllable} %in the time interval $I$ 
if for each state ${\bf z}$ 
there are an instant $t>0$ and an input function $u(.)$, defined in $[0, \, t]$,  such that the solution ${\bf z}(.)$ to equation  (\ref{eqnarray:sy1}) satisfies
${\bf z}(0)={\bf z}$, $\;{\bf z}(t)={\bf 0}$; 

$(C)\;$ {\it completely observable} if every initial state ${\bf z}(0)$ can be determined by observing the future outputs ${\bf y}(s)$  for   $s \in [0, \tau]$  for some $\tau > 0$;

$(D)\;$  {\it completely re-construictible} if every actual state ${\bf z}(t)$ can be determined by observing the past outputs ${\bf y}(s)$ for   $s \in [t - \tau, t]$  for some $\tau > 0$.

By the linearity, for  any continuous system  (\ref{eqnarray:sy1})-(\ref{eqnarray:sy2}),

$(E)$  {\it the controllability space and reachable space coincide and do not depend on the lenght $t$ of the time interval;}  
$(F)$  {\it the observability condition and re-construictibility condition are equivalent and do not depend on the lenght $\tau$ of the time interval.}

Here we consider strict casual systems with one scalar input and one scalar output;  hence
\begin{equation}    \label{equation:zerosylin1}
 {\bf d}={\bf O}\,  ,   \qquad  m=1=p \, ,
\end{equation}
${\bf g}\,$ is a column vector of dimension $n$,  $\,{\bf h}\,$ is a row vector of dimension $n$,  
and  equations (\ref{eqnarray:sy1}), (\ref{eqnarray:sy2}) become
%%%%%%%%%%%%%%%%%%%%%%%%%%%%%%%%%%%
\begin{equation} 																														%		 \label{equation:sys}
\dot{\bf z} ={\bf F}\,{\bf z}+{\bf g}\,u  \; ,           \label{equation:uno}
\end{equation}
\begin{equation}
y ={\bf h} \cdot {\bf z} \; .                                   \label{equation:sylin2}
\end{equation}
%%%%%%%%%%%%%%%%%%%%%%%%%%%%%%%%%%%%%

We use the following theorem, which holds for system (\ref{equation:uno})-(\ref{equation:sylin2} ),  
with one scalar control input and one scalar output (see \cite{FM:Appunti}). 

\begin{theorem}                                   \label{th:uno}
The mechanical system $\Sigma$, represented by (\ref{equation:uno})-(\ref{equation:sylin2}), is completely reachable and completely observable if and only if the polynomials
\begin{equation}
det(z{\bf I}-{\bf F}) \quad  {\rm and}  \quad  {\bf h}\,adj(z{\bf I}-{\bf F})\,{\bf g}
\end{equation}
have no common root.
\end{theorem}
%%%%%%%%%%%%%%%%%%%%%%%%%%%%%
%%%%%%%%%%%%%%%%%%%%%%%%%%%%%%

\subsection{Mechanics of spring-mass chains with viscous damping}                               \label{subsection:msmch}
%%%%%

%\begin{figure}[h]
%\centerline{\epsfig{figure=molle1W.eps,width=10cm}}
%\end{figure}

%\end{sidewaysfigure}
%\end{sidewaysfigure}

\begin{figure}   
\centerline{\epsfbox{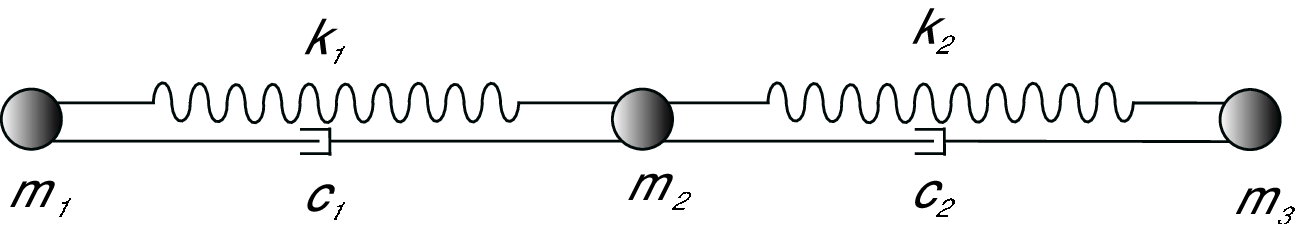}}
\caption{Physical model of system $\;\Sigma_2$.}
\label{landfig:uno}
\end{figure}

\begin{figure}
\centerline{\epsfbox{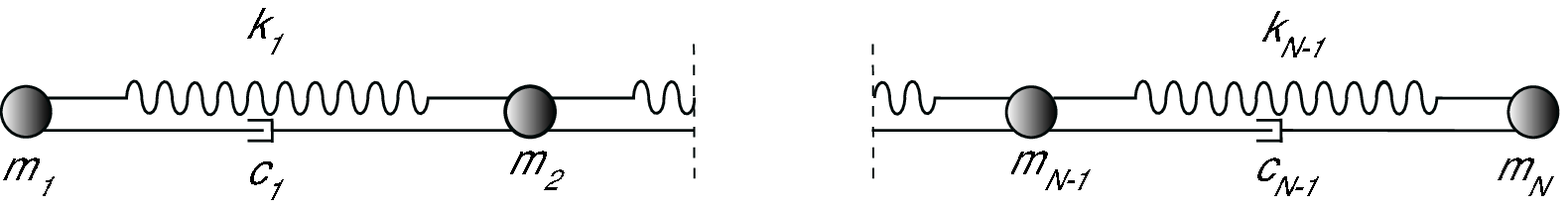}}
\caption{Physical model of system  $\;\Sigma_N$,  $\; N > 2$.}
\label{landfig:due}
\end{figure}

We consider the chain of $N \geq 2$  mass points $P_1, \, P_2, \ldots , P_N$ of respective masses $m_1, \, m_2, \ldots , m_N$. Each point interacts only with its nearest neighbours since any two consecutive points $P_i , \, P_{i+1}$ are linked by a linear spring having elastic constant  $k_i>0$ and natural lenght  $\ell_i$. 
Moreover, there is viscous damping since  consecutive mass points are connected by dashpots of constants  $c_i \geq 0$ (see Figure \ref{landfig:uno} and Figure \ref{landfig:due}).
We denote $z_i$ the displacement of the point $P_i$ from its equilibrium position, so that $z_i=0$  for each $i=1$ to $N$  corresponds to no deformations in the springs.  Of course, this system has $N$ degrees of freedom and the $z_i$ are free coordinates for it.

The equations of motion of the mass points $P_i$ have the form
%%%%%%%%%%%%%%%%%%%%%%%%%%%%%%%%%%%
\begin{equation}  														         						 
m_1 {{d^2}\over{dt^2}}z_1 = k_1 (z_{2}-z_1) + c_1 (\dot{z}_{2}-\dot{z}_1) + f^{ext}_1    \; ,   \label{eq:eqm1}  
\end{equation}
\begin{eqnarray}   m_i {{d^2}\over{dt^2}}z_i \,=\, k_{i-1} (z_{i-1}-z_i)+ k_i (z_{i+1}-z_i)  \\  \nonumber
+ c_{i-1}(\dot{z}_{i-1}-\dot{z}_i)+c_i(\dot{z}_{i+1}-\dot{z}_i)  \qquad( i=2,\ldots,N-1) \, ,  \label{eqnarray:eqmi}
 \end{eqnarray}
 \begin{equation} 
m_N {{d^2}\over{dt^2}}z_N = k_{N-1}(z_{N-1}-z_N)+ c_{N-1}(\dot{z}_{N-1}-\dot{z}_N) + f^{ext}_N  \; .   \label{eq:eqmn}
\end{equation}
%%%%%%%%%%%%%%%%%%%%%%%%%%%%%%%%%%%
Here we  study 

$\,(i)\,$ the controllability of the state system by the input control $u=f^{ext}_1$, the external force applied to the first mass point $P_1$,  
and 

$\,(ii)\,$ the observability of each  state $\,{\bf z}\,$ by using as output  the position $y=z_N$ of the last mass point. 
%
% \item[(C)] {\it completely observable} if every initial state ${\bf z}(0)$ can be determined by observing the future outputs % ${\bf y}(s)$  for   $s \in [0, \tau]$  for some $\tau > 0$.???

Hence we put $\,f^{ext}_N=0\,$ and by introducing the variables $z_{N+i} =  {{d}\over{dt}}z_i$, equations  (\ref{eq:eqm1})-(\ref{eq:eqmn}) give rise to the first order system of $2N$ linear equations
%%%%%%%%%%%%%%%%%%%%%%%%%%%%%%%%%%%%%
%%%%%%%%%%%%%%%%%%%%%%%%%%%%%%%%%%%
\begin{equation}  											 \label{eq:eq11b}			         						 
{{d}\over{dt}}z_i = z_{N+i}   \; ,  \quad \quad   i=1,\ldots, N  ,                           
\end{equation}
\begin{equation}  \label{eq:eq11b2}
m_1 {{d}\over{dt}}z_{N+1} = k_1 (z_{2}-z_1)  + c_1 (z_{N+2}-z_{N+1}) + u  ,  
\end{equation}
\begin{eqnarray}                                                   \label{eqnarray:eq11b3}  \nonumber
m_i {{d}\over{dt}}z_{N+i}\, =\, k_{i-1} (z_{i-1}-z_i)+ k_i (z_{i+1}-z_i) + \\
c_{i-1}(z_{N+i-1}-z_{N+i}) + c_i(z_{N+i+1}-z_{N+i})    \qquad (i=2,\ldots, N-1)
 \end{eqnarray}
\begin{equation}          \label{eq:eq11b4}  
m_N {{d}\over{dt}} z_{2N}  = k_{N-1}(z_{N-1}-z_N) + c_{N-1}(z_{2N-1}-z_{2N})  \; .
\end{equation}

In order to write equations (\ref{eq:eq11b}) to (\ref{eq:eq11b4})  in matrix form, we put
%%%%%%%%%%%%%             %%%%%%%%%%%%%%%%%%%%%%
\begin{eqnarray}  										 \label{eq:abcde} 			\nonumber	         						 
 {\bf X}(x_1, \,x_2, \cdots , \, x_{N-1})
 = \left( \begin{array}{cccccc}
{{x_1}\over{m_1}} & -{{x_1}\over{m_1}} & 0 & .& . & 0                                                         \\
-{{x_1}\over{m_2}} & {{x_1+x_2}\over{m_2}} & -{{x_2}\over{m_2}}  & 0 & .  &  0                               \\
0 & -{{x_2}\over{m_3}} & {{x_2+x_3}\over{m_3}} &  -{{x_3}\over{m_3}}  & 0 & .                             \\
0 & 0 & . &  .  & . & .                                                                                    \\
. & . & . &  .  & -{{x_{N-2}}\over{m_{N-2}}}  & 0                                                          \\
.& . & 0 & -{{x_{N-2}}\over{m_{N-1}}} & {{x_{N-2}+x_{N-1}}\over{m_{N-1}}} & -{{x_{N-1}}\over{m_{N-1}}}  \\
0 & . & 0 & 0 & -{{x_{N-1}}\over{m_N}} & {{x_{N-1}}\over{m_N}}  \\
\end{array}        \right)
\end{eqnarray}
%%%%%%%%%%%%%%%%%%%%

%%%%%%%%%%%%%             %%%%%%%%%%%%%%%%%%%%%%
\begin{equation}  													 \label{eq:abcde2} 	         						 
 {\bf K}_N={\bf X}(k_1, \,k_2, \,\cdots , k_{N-1})\, ,  \qquad   {\bf C}_N={\bf X}(c_1, \,c_2, \,\cdots , c_{N-1})\, , 
\end{equation}
%%%%%%%%%%
%%%%%%%%%%          
and
%%%%%%%%%%%%%%%%%%%%%%
\begin{eqnarray}                                                  \label{eqnarray:eq11bcd}	       						 
 {\bf F}_{N}= \left[ \begin{array}{cccc}
{\bf O}_N & {\bf I}_N   \\
-{\bf K}_N & -{\bf C}_N  \\
\end{array}        \right] 
\end{eqnarray}
%%%%%%%%%%%%%%%%%%%%
where $\;{\bf O}_N\,$ is the square zero matrix of order $\,N\,$
and $\;{\bf I}_N\,$ is the identity matrix of order $\,N\,$;  moreover we put
%%%%%%%%%%%%%%%%%%%%%%%%%%%%%%%%%%%%%%%%%%%%%%%%%%%%%%%%%%%%%%%%%%%%%%%%%%%%%%%%%
\begin{eqnarray}  		
 {\bf z} =
 \left( \begin{array}{c}
z_1  \\
z_2  \\
.  \\
.  \\
.  \\
z_N  \\
z_{N+1}  \\
.\\
.\\
z_{2N} \\
\end{array}   
    \right)    \,  ,  \quad
{\bf g}= \left( \begin{array}{c}
0  \\
.  \\
.  \\
.  \\
0  \\
1/m_1  \\  
0 \\
.  \\
.\\
.  \\
0 \\
\end{array}       
 \right)
 \;  ,   \quad 
  {\bf h}= \left( \begin{array}{ccccccc}
0 & \ldots & 0 & 1 & 0 & \ldots & 0 
\end{array}       \right) \end{eqnarray}
%%%%%%%%%
where $1/m_1$ is in $(N+1)$th  place of column {\bf g}, of dimension $2N$, and $1$ is in $N$th  place of row {\bf h}, of dimension $2N$.

%%%%%%%%%%%%%%%%%%%%%%%%%%%%%%%%%%%%%%%%%%%%%%%%%%%%%%%%%%%%%%%%%%%%%%%%%%%%%%%%%%

Hence equations (\ref{eq:eq11b})-(\ref{eq:eq11b4}) in matrix form rewrite as:
%%%%%%%%%%%%%%%%%%%%%%%%%%%%%%%%%%%
\begin{eqnarray}  																														%		 \label{eqnarray:sys}
\dot{\bf z} ={\bf F}\,{\bf z}+{\bf g}\,u  \; ,       \\    \label{eqnarray:sylinx} %\label{msmch}
y ={\bf h} \,\cdot \, {\bf z} \; .                                          \label{eqnarray:sylinx2}
\end{eqnarray}
%%%%%%%%%%%%%%%%%%%%%%%%%%%%%%%%%%%%%

%%%%%%%%%%%%%%%%%%%%%%%%%%%%%%%%%%%%%

\section{Two useful lemmas}                      \label{section:two}

For any square matrix $\;A\;$  we use the notation
\begin{equation}                                     \label{eq:not}
\left\| A  \right\| = Det A \;  .
\end{equation}

In order to prove the two key theorems of the present paper, {\it i.e.}  Theorem \ref{th:adjpol} and  Theorem \ref{th:tcp}, we need the two lemmas below.
\begin{lemma}  \label{lemma:uno} 
 For any choice of the positive integers $\, \alpha$,  $\,\beta$ and of the real number $\, \rho$, let
 
$ O_{\alpha \times \beta}$ be the $\alpha \times \beta$ zero matrix,

$I_\alpha$ be the identity matrix of order $\alpha$,

$B_\beta$ be any matrix of order $\beta$, and

%%%%%%%%%%%%%             %%%%%%%%%%%%%%%%%%%%%%
\begin{eqnarray}  											       \label{eqnarray:AOB}  			         						 
A = \left[ \begin{array}{cc}
O_{\alpha \times \beta}  & \rho\,I_\alpha   \\
B_\beta  & O_{\beta \times \alpha}     \\
\end{array}        \right] \;  .
\end{eqnarray}
%%%%%%%%%%%%%%%%%%%%%%%%%%%%%%%%%%%%%%%%%%%%%%%%%%%%%%%%%%%%%%%%%%%%%%%%%%%%%%%%%%%%%%%%%%%%%%%%%%%%%
\medskip

Then
%%%%%%%%%%%%%             %%%%%%%%%%%%%%%%%%%%%%
\begin{equation}  														         		 \label{eq:DetAb}  					 
\left\| A  \right\| = \rho^{\alpha}(-1)^{\alpha(\beta+2)}\, \left\| B_\beta \right\|  \quad.
\end{equation}
%%%%%%%%%%%%%%%%%%%%%%%%%%%%%%%%%%%%%%%%%%%%%%%%%%%%%%%%%%%%%%%%%%%%%%%%%%%%%%%%%%%%%%%%%%%%%%%%%%%%%
\end{lemma}

%\end{flushright}

\underline{Proof.}$\;$  By expanding $A$ in the cofactors of the $1st$ row yields 
\begin{equation}  			                         											         		 \label{eq:ddaabb}  					 
 \left\| A \right\|= \rho (-1)^{\beta+2}\,
  \left \|  \begin{array}{cc}
           O_{(\alpha-1)\times \beta} & \rho\,I_{\alpha-1} \\
                             	B_\beta &  O_{\beta \times (\alpha-1)} \\
				 \end{array} \right \| \; . 
\end{equation}

Repeating this step to the matrix
\begin{equation}  			                         											         		 \label{eq:ddaabbnnn}  					 
  \left \|  \begin{array}{cc}
           O_{(\alpha-1)\times \beta} & \rho\,I_{\alpha-1} \\
                             	B_\beta &  O_{\beta \times (\alpha-1)} \\
				 \end{array} \right \| \; ,
\end{equation}

consecutively for each $i=2, \dots , \,\alpha$,  yields
\begin{equation}  			                         											         		 \label{eq:ddaaiiii}  					
  \left\| A \right\|= \rho^i (-1)^{i \times (\beta+2)}\,
  \left \|  \begin{array}{cc}
           O_{(\alpha -i)\times \beta} & \rho\,I_{\alpha-i} \\
                             	B_\beta &  O_{\beta \times (\alpha-i)} \\
				 \end{array} \right \| \; . 
\end{equation}
and thus for $i=\alpha$ we obtain equality (\ref{eq:DetAb}).    %$\quad \diamondsuit$
\begin{flushright}
Q.E.D.
\end{flushright}

\bigskip

%\textbf{5. The proof of the second theorem}
%\vspace{1cm}

\begin{lemma}                                     \label{lemma:tre} 
Let $\gamma$ be a positive integer and let $A$ be a square matrix of order $2\gamma-1$  having the form

\begin{equation}                              \label{eq:A}  
A= \left[ \begin{array}{ccc}
        z\,I_{\gamma-1}  &  -I_\gamma \\
        Row_{\gamma-1} (0)  \\
        UpTr (a_1, a_2, \ldots, a_{\gamma-1})  &  
        UpTr (b_1, b_2, \ldots, b_{\gamma-1})  \; Col_{\gamma-1}(\#)
\end{array} \right] ,
\end{equation}               

where $\,(i)\; Row_{\gamma-1} (0)\,$  is the row  of zeros of dimension $\,\gamma-1$,  
$\,(ii)\; UpTr (x_1, x_2, \ldots, x_{\gamma-1})\,$ denotes any upper triangular matrix having the numbers
$\,x_1, x_2, \ldots, x_{\gamma-1}\,$ in the main diagonal and
 $\,(iii)\; Col_{\gamma-1}(\#)\,$ denotes any column of dimension $\,\gamma-1$.
Then

\begin{equation}                              \label{eq:Agamma}  
\left\| A \right\| = (-1)^\gamma (a_1+z b_1) (a_2+z b_2) \ldots (a_{\gamma-1}+z b_{\gamma-1})\; .
\end{equation}               

\end{lemma}
% ??????????????????????????????????????????????????????????????????????????????????????????????????????????????''''
\underline{Proof.}$\;$  For each  $\,i=1,  \ldots, N-1 \,$  multiply the $\,(N-1+i)th\,$ column of $\,A\,$ by $\,z\,$ and add the result to the $\,ith\,$ column;  we obtain a matrix $B$ having the form
%%%%%%%%%%%%%%%%%%%%%%%%%%%%%%%%%%%%%%%%%%%%%%%%%%%%%%%%%%%%%%%%%%%%%%%%%%%%%%%%%%%%%%%%%%%%%%%%%%%
\begin{equation}                              \label{eq:B}  
B= \left[ \begin{array}{ccc}
        O_{\gamma \times (\gamma-1)}  &  -I_\gamma \\
       UpTr(a_j+z b_j\,: \, 1\leq j \leq \gamma-1)  &  UpTr (b_j\,: \, 1\leq j \leq \gamma-1)   \;\, Col_{\gamma-1}(\#)
\end{array} \right] .
\end{equation}               
       % UpTr (a_1+z b_1, a_2+z b_2, \ldots, a_{\gamma-1}+z b_{\gamma-1})  &  UpTr (b_1, b_2, \ldots, b_{\gamma-1})  \;  
       
Now for $\,i=1,  \ldots, \gamma \,$,  multiplying the $\,ith\,$ row of $\,B\,$ by suitable quantities 
and adding the results to the $\,(N+1+i)th\,$ row, one obtains a matrix $B'$ having the form
%%%%%%%%%%%%%%%%%%%%%%%%%%%%%%%%%%%%%%%%%%%%%%%%%%%%%%%%%%%%%%%%%%%%%%%%%%%%%%%%%%%%%%%%%%%%%%%%%%%
\begin{equation}                              \label{eq:BB}  
B' = \left[ \begin{array}{ccc}
        O_{\gamma \times (\gamma-1)}  &  -I_\gamma \\
        %Row_{\gamma-1} (0)  \\
 UpTr(a_j+z b_j\,: \, 1\leq j \leq \gamma-1) &  O_{(\gamma-1) \times \gamma}
\end{array} \right] .
\end{equation}               

By the properties of determinants we have   $\,\left\|A\right\|=\left\|B\right\|=\left\|B'\right\|\,$.
By Lemma \ref{lemma:uno}  with  $\, \rho = -1 \,$  we have
 
$$\left\|B'\right\|=(-1)^{\gamma}(-1)^{\gamma(\gamma+1)} \left\|UpTr(a_j+z b_j\,: \, 1\leq j \leq \gamma-1)\right\|\, ;$$

thus, since $\gamma(\gamma+1)$ is even,
$$\left\|A\right\|=\left\|B'\right\|=(-1)^{\gamma} \left\|UpTr(a_j+z b_j\,: \, 1\leq j \leq \gamma-1)\right\|\, .$$
Now  equality  (\ref{eq:Agamma})  holds since the determinant of a upper triangular matrix equals the product of its diagonal elements.

\begin{flushright}
Q.E.D.
\end{flushright}

%%%%%%%%%%%%%%%%%%%%%%%%%%%%%%%%%%%%%%%%%%%%%%%%%%%%%%%%%%%%%%%%%%%%%%%%%%%%%%%%%%%%%%%%%%%%%%%%%%%%%%%%%%%%%%%%%%%%%%
%%%%%%%%%%%%%%%%%%%%%%%%%%%%%%%%%%%%%%%%%%%%%%%%%%%%%%%%%%%%%%%%%%%%%%%%%%%%%%%%%%%%%%%%%%%%%%%%%%%%%%%%%%%%%%%%%%%%%%

\section{Form of the adjoint polynomial}                               \label{adj} 

\begin{theorem}                                     \label{th:adjpol}
In connection with any  positive integer $\,N\geq 2\,$ consider the square matrix $\,F_N\,$ of order $2N$ defined by
(\ref{eq:abcde})-(\ref{eqnarray:eq11bcd});  moreover
%\vspace{.5cm}

 $(i)\;$ let $\; a_N = zI_{2N}-F_N \,$, thus
 \begin{equation}                                     \label{eq:matrixaN}
 a_N = \left[ \begin{array}{cc}
       z\,I_N  &  -I_N \\
       K_N  &  C_N+z\,I_N
  \end{array} \right] \; ;
 \end{equation}                                      

 $(ii)\;$   let 
 \begin{equation}                                     \label{eq:hhh} 
h_N=
\left [ \begin{array}{ccccccc}
	0 & \dots & 0 & 1 & 0 &   \dots &  0 \\
\end{array} \right ]
\end{equation}
be the row of dimension $\,2N$, with components $\, \delta^{Ni}$,  $ \; i=1, 2, \dots , 2N$ ($\delta=$ Kronecker's delta);
%having $\,1\,$  in the  $N$th place and $0$ otherwise, {\it i.e.},  $h_N^i=\delta^{Ni}, \; i=1, 2, \dots , 2N$;

 $(iii)\;$  let 
 \begin{equation}                                     \label{eq:hhhbbb} 
g_N=
\left [ \begin{array}{ccccccc}
	0 & \dots & 0 & \frac{1}{m_1} & 0 & \dots &  0 \\
\end{array} \right ]^T
\end{equation}
be the column of dimension $\,2N$, with components $\, \frac{1}{m_1}\delta^{(N+1)i}$,  $ \; i=1, 2, \dots , 2N$.

Then 
 \begin{equation}                                     \label{eq:hhhbxx} 
h_N\,adj(a_N)\,g_N\,=\,\frac{(k_1+zc_1)\,(k_2+zc_2)\,\ldots \, (k_{N-1}+zc_{N-1})}{m_1\,m_2\,\ldots \,m_N} \; .
\end{equation}

\end{theorem}   

\underline{Proof.} $\;$ The polynomial $\,h_N\,adj(a_N)\,g_N\,$ equals 
$\,c^{N\; N+1}/m_1\,$,  where  $\,c^{N\; N+1}\,$ is the $\,(N, \, N+1)\,$ component of  the matrix $\,c=adj(a_N)$;
hence we have  

\begin{equation}                                     \label{eq:nowwe} 
m_1\,h_N\,adj(a_N)\,g_N\,=\,c^{N\; N+1}=(-1)^{N+(N+1)}\left\|A_{N+1\; N}\right\|=-\left\|A_{N+1\; N}\right\| \; ,
\end{equation}
  
where $\,A_{N+1\; N}\,$ is the submatrix of $\,a_N\,$ obtained by deleting the $(N+1)$th row and the $N$th column.
Now we observe that  $\,A_{N+1\; N}\,$ has the form (\ref{eq:A}), that is,
\begin{equation}                                     \label{eq:adadad} 
A_{N+1\; N}=
   \left[ \begin{array}{cc}
       z\,I_{N-1}  &  -I_N  \\
       Row_{N-1}(0)  &     \\
       UpTr(a_1,\,a_2,\dots , a_{N-1}) &  UpTr(b_1,\,b_2,\dots , b_{N-1}) \; Col_{N-1}(\#)                                  
               \end{array} \right] \, , 
\end{equation}
with 
\begin{equation}                                     \label{eq:where} 
a_1 =-\frac{k_1}{m_2}\, ,  \quad a_2 =-\frac{k_2}{m_3}\, , \quad \dots \; ,
\quad  a_{N-1} =-\frac{k_{N-1}}{m_{N}}\;,
\end{equation}

\begin{equation}  
b_1 =-\frac{c_1}{m_2}\, ,  \quad b_2 =-\frac{c_2}{m_3}\, , \quad \dots \; ,
\quad  b_{N-1} =-\frac{c_{N-1}}{m_{N}}\;,
\end{equation}
and  $\,  Col_{N-1}(\#)  \,$ is any column of dimension $\,N-1$.
Now Lemma \ref{lemma:tre} yields
\begin{equation}                                     \label{eq:where2} 
\left\| A_{N+1\; N} \right\|  = (-1)^N (a_1+z b_1) (a_2+z b_2) \ldots (a_{N-1}+z b_{N-1})\; ,
\end{equation}

and thus equalities (\ref{eq:nowwe}) to (\ref{eq:where2}) yield    
$$h_N\,adj(a_N)\,g_N\, = \,\frac{c^{N\; N+1}}{m_1}=(-1)^{N+1} 
                           \frac{(a_1+z b_1) (a_2+z b_2) \ldots (a_{N-1}+z b_{N-1})}{m_1}$$
$$\,=\,(-1)^{N+1+N-1} \frac{(k_1+z c_1) (k_2+z c_2) \ldots (k_{N-1}+z c_{N-1})}{m_1\,m_2\,\dots \,m_N}\,.$$

\begin{flushright}
Q.E.D.
\end{flushright}
%%%%%%%%%%%%%%%%%%%%%%%%%%%%%%%%%%%%%%%%%%%%%%%%%%%%%%%%%%%%%%%%%%%%%%%%%%%%%%%%%%%%%%%%%%%%%%%%%%%%%%%%%%%%5

\section{Form of the characteristic polynomial}                               \label{cp} 
Let us remind that the characteristic polynomial $P_N(z)$ of $F_N$ is defined by

\begin{equation}                                     \label{eq:4} 
P_N(z)=\left\| a_N \right\| \, , \quad {\rm with} \quad a_N = zI_{2N}-F_N \; . 
\end{equation}
%\vspace{.5cm}
The next theorem characterizes the characteristic polynomials of the mechanical systems considered here.

Note that the chain of mass points which is considered here has no springs connecting the first point and the latter to a fixed external wall. 

For this reason the characteristic polynomial has the twice root zero.

\begin{theorem}                                     \label{th:tcp} 
In connection with any  positive integer $\,N\,$ consider the square matrix $\,F_N\,$ of order $2N$ defined by
(\ref{eq:abcde})-(\ref{eqnarray:eq11bcd}), and put 

\begin{equation}                                    \label{th:bck} 
	b_N=zc_N + k_N \, .
\end{equation}

Then the equalities  
\begin{equation} \label{eq:3-1} 
P_1(z)=z^2 \; ,  \qquad \quad D_0=1  \; ,
\end{equation}
%\vspace{.5cm}
\begin{equation} \label{eq:3-2} 
D_N(z)=P_N(z) + \frac{b_N}{m_N} D_{N-1}(z)   \; , \quad  N=1,2,\ldots \; , %  {\rm for}
\end{equation}
%\vspace{.5cm}
\begin{equation} \label{eq:3-3} 
P_N(z)=( z^2 +\frac{b_{N-1}}{m_N}) P_{N-1} (z) +z^2 \frac {b_{N-1}}{m_{N-1}} D_{N-2}(z)   \;  , \quad  N=2,3 \ldots \, ,   
\end{equation}
inductively define the sequence $\, \left\{ P_N(z)  \right\}_{N \geq 1} \,$  of the characteristic polynomials of $\,F_N$.

Moreover, let 
\begin{equation}                                     \label{eq:3-4} 
\{1\}_N = z^2 I_N + z C_N + K_N +  W_N   \, ,  
\end{equation}
with 
%%%%%%%%%%%%%%%%%%%%%%%%%%%%%%%%%%%%%%%%%%%%%%%%%%%%%%%%%%%%%%%%%%%%%%%%%%%%%%%%%%%%%%%%%%%%%%%%%%%%%%%%%%%%
\begin{equation}  			                         											         		 \label{eq:wwwww}  					 
 W_N \,=\, \left[  \begin{array}{cc}
           O_{(N-1)} & Col_{(N-1)}(0)  \\
           Row_{(N-1)}(0) & (z\frac {c_N}{m_{N}}+\frac {k_N}{m_{N}}) \\
				 \end{array} \right]\; . 
\end{equation}
%%%%%%%%%%%%%%%%%%%%%%%%%%%%%%%%%%%%%%%%%%%%%%%%%%%%%%%%%%%%%%%%%%%%%%%%%%%%%%%%%%%%%%%%%%%%%%%%%%%%%%%%%
Then the quantities $\,D_k\, $ in (\ref{eq:3-1})-(\ref{eq:3-3}) are the determinants of  matrices $\,\{1\}_k \,$:
\begin{equation}     \label{eq:3-4nn} 
\quad D_k=\left\|{\{1\}_k}\right\| \; .
\end{equation}

 \end{theorem}   

\underline{Proof.}$\quad$  
We prove the theorem by induction.
Hence  we firstly note that the inductive formulae (\ref{eq:3-1})-(\ref{eq:3-3}) hold for $N=2$; in fact, see (\ref{eqnarray:eq11bcd}), for
\begin{equation}                                     \label{eq:44} 
F_2=
\left [ \begin{array}{cccc}
	 O_2  &  I_2  \\
	-K_2  &  -C_2
\end{array} \right ]	\; ,
\end{equation}
by (\ref{eq:4})  
we have 
\begin{equation}                                     \label{eq:4444} 
a_2=
\left [ \begin{array}{cccc}
	 zI_2  &  -I_2  \\
	 K_2  &  (zI_2+C_2)
\end{array} \right ]	
\end{equation}

and
\vspace{.5cm}
$$P_2(z)=\left\|a_2\right\|=
\left\| \begin{array}{cc}
	(z^2+z\frac{c_1}{m_1}+\frac{k_1}{m_1}) &  (-z\frac{c_1}{m_1}-\frac{k_1}{m_1}) \\
	(-z\frac{c_1}{m_2}-\frac{k_1}{m_2}) & (z^2+z\frac{c_1}{m_2}+\frac{k_1}{m_2})
\end{array} \right\| $$
\begin{equation}                                     \label{eq:xxx} 
=\left(\frac{k_1}{m_1}+\frac{k_1}{m_2}\right)z^2+\left(\frac{c_1}{m_1}+\frac{c_1}{m_2}\right)z^3+z^4 \; .
\end{equation}
\vspace{.5cm}

It is easy to verify that using (\ref{eq:3-1})-(\ref{eq:3-3}) with $N=2$ one arrives at the same result.
That is, the expression of $P_2(z)$ given by its definition (\ref{eq:4}) coincides with the expression given by equalities (\ref{eq:3-1})-(\ref{eq:3-3}).

Now, in order to use the induction principle, assume the inductive hypothesis: let (\ref{eq:3-1})-(\ref{eq:3-3}) hold for an arbitrarly given integer $N$; then  consider
%\vspace{.5cm}
\begin{equation}                                     \label{eq:y} 
F_{N+1} =
\left [ \begin{array}{cc}
	O_{N+1}  & I_{N+1}\\
	-K_{N+1} & -C_{N+1}
\end{array} \right ] \; ,	
\end{equation}
with $\,K_{N+1}\,$ and $\,C_{N+1}\,$ defined accordingly with (\ref{eq:abcde}), (\ref{eq:abcde2}).

By  (\ref{eq:4}) %and (\ref{eq:5}) 
we have 
\begin{equation}                                     \label{eq:6} 
P_{N+1}(z)=\left\|a_{N+1}\right\|, \hspace{.5 cm} 
a_{N+1}=
\left [ \begin{array}{cc}
	z I_{N+1} &  -I_{N+1}\\
	K_{N+1} & (z I_{N+1}+C_{N+1})
\end{array} \right ] 
\end{equation}     
%%%   \vspace{.5cm}

Now, for each $\, i=1, \ldots, N$, we multiply the $(N+1+i)$th column of the matrix $a_{N+1}$ by $z$ and add the result to the $ith$-column; we obtain the matrix 
\begin{equation}                                     \label{eq:6bis} 
\left [ \begin{array}{cc}
	O_{N+1}  &  I_{N+1}\\
	(z^2 I_{N+1}+zC_{N+1}+K_{N+1}) & (z I_{N+1}+C_{N+1}) 
\end{array} \right ]	 \, .
\end{equation}                                   
%%  \vspace{.5cm}

Hence for each $\,i=1, \ldots, N+1\,$  multiplying the $i$th row of the latter matrix by suitable quantities and adding the results to the $\,(N+1+i)$th row, we obtain the matrix

\begin{equation}                                     \label{eq:7} 
a'_{N+1} =
\left [ \begin{array}{cc}
	O_{N+1}  & -I_{N+1}\\
	(z^2 I_{N+1}+zC_{N+1}+K_{N+1}) &  O_{N+1}
\end{array} \right ]	 \, .
\end{equation}  
%\vspace{.5cm}
Of course, by the determinant properties,
%\vspace{.5cm}

\begin{equation}                                     \label{eq:8} 
\left\|a'_{N+1}\right\| = \left\|a_{N+1}\right\| = P_{N+1}(z) \; . 
\end{equation} 

By  (\ref{eq:7})-(\ref{eq:8}) Lemma \ref{lemma:uno}, for $\, \alpha=\beta=N+1 \,$ and $\, A=a_{N+1}$,  yields

\begin{equation}                                     \label{eq:9} 
\left\|a'_{N+1}\right\| = (-1)^{N+1} (-1)^{(N+1)(N+3)} \left\|z^2 I_{N+1}+zC_{N+1} + K_{N+1}\right\| = P_{N+1}(z)\;.
\end{equation} 

Since  $\,N+1+(N+1)(N+3)=(N+1)(N+4)\,$ is even for each $N$,  we have
$(-1)^{N+1} (-1)^{(N+1)(N+3)} = 1$,  and thus
\begin{equation}                                     \label{eq:10} 
\left\|z^2 I_{N+1}+zC_{N+1} + K_{N+1}\right\| = P_{N+1} (z)
\end{equation}

Now, by expanding the matrix
$\,z^2 I_{N+1} + zC_{N+1} + K_{N+1} \,$
in the cofactors of its last row, we obtain
$$P_{N+1}(z) = \|z^2 I_{N+1} + zC_{N+1} + K_{N+1}\| $$ 
 \begin{equation}                                     \label{eq:111} 
 =(-1)^{2N+2} \left(z^2 + z\frac {c_N}{m_{N+1}} + \frac {k_N}{m_{N+1}}\right) \left\|\left\{1\right\}_N \right\| + 
  (-1)^{2N+1} \left(- z\frac {c_N}{m_{N+1}}- \frac {k_N}{m_{N+1}}\right) \left\|\left\{2\right\}_N \right\|   \, ;       
  \end{equation} 
thus,
\begin{equation}                                     \label{eq:12} 
P_{N+1} (z) = (z^2 + z\frac {c_N}{m_{N+1}}+ \frac{k_N}{m_{N+1}}) \|\{1\}_N \| +
(z\frac {c_N}{m_{N+1}}+\frac {k_N}{m_{N+1}}) \left\|\left\{2\right\}_N \right\| \; ,
\end{equation}
where,
for
\begin{equation}
	\alpha_j\,=\,z^2 + z\frac{c_j + c_{j+1}}{m_j}+ \frac{k_j + k_{j+1}}{m_j} \; ,
\end{equation}

 the matrices $\,\{1\}_N$, $\,\{2\}_N \,$ % accordingly with (\ref{eq:3-4}),  
are defined by

%%%%%%%%%%%%%%%%%%%%%%%%%%%%%%%%%%%%%%%%%%%%%%%%%%%%%%%%%%%%%%%%%%%%%%%%%%%%%%%%%%%%%%%%%%%%%%%%%%%
\begin{equation}                                     \label{eq:13ZZqua} 
\,\{1\}_N \, = \, \left[ \begin{array}{l}
                         Z_1  Z_2 
                          \end{array}\right]  \; ,
                          \qquad  \,\{2\}_N \, = \, \left[ \begin{array}{l}
                         Z_1  Z_3 
                          \end{array}\right] 
\end{equation}
  
%%%%%%%%%%%%%%%%%%%%%%%%%%%%%%%%%%%%%%%%%%%%%%%%%%%%%%%%%%%%%%%%%%%%%%%%%%%%%%%%%%%%%%%%%%%%%%%%%%%%%%
%  GENERICO  2N
%%%%%%%%%%%%%%%%%%%%%%%%%%%%%%%%%%%%%%%%%%%%%%%%%%%%%%%%%%%%%%%%%%%%%%%%%%%%%%%%%%%%%%%%%%%%%%%%%%%%%

with $\,Z_1$,  $\,Z_2$  and $\,Z_3\,$   respectively given by
%%%%%%%%%%%%%%%%%
\begin{equation}                                     \label{eq:13Z1} 
\left[ \begin{array}{cccc}
   (z^2 +z\frac{c_1}{m_1}+\frac{k_1}{m_1}) &  (-z\frac{c_1}{m_1}- \frac{k_1}{m_1}) & 0 & \dots      \\
    (-z\frac{c_1}{m_2}-\frac{k_1}{m_2})  &  \alpha_1  & (-z\frac{c_2}{m_2}-\frac{k_2}{m_2})   & \dots \\
     0  &   (- z\frac{c_2}{m_3}- \frac{k_2}{m_3})  &  \alpha_2  & \dots  \\
     0  &   0  &  (- z\frac{c_3}{m_4} - \frac{k_3}{m_4})  & \dots   \\
      0  &   0  &   0   & \dots   \\
      0  &   0  &   0   & \dots   \\
     \dots  &   \dots   &  \dots   & \dots   \\
\end{array}\right] 
\end{equation}  
%%%%%%%%%%%%%%%%%%
%\vspace{.4cm}
%%%%%%%%%%%%%%%%%
\begin{equation}                                     \label{eq:13Z222} 
\left[ \begin{array}{cccc}
  \dots  &   \dots   &  \dots   & \dots   \\
   \dots &   0  &   0  &   0    \\
   \dots  &   0  &   0  &   0    \\
   \dots  &  (- z\frac{c_{N-3}}{m_{N-3}} - \frac{k_{N-3}}{m_{N-3}})  &   0  &  0  \\
   \dots  &     \alpha_{N-3}    &  (-z\frac{c_{N-2}}{m_{N-2}}- \frac{k_{N-2}}{m_{N-2}}) & 0       \\
   \dots  & (-z\frac{c_{N-2}}{m_{N-1}}-\frac{k_{N-2}}{m_{N-1}})  &  \alpha_{N-2}  & (-z\frac{c_{N-1}}{m_{N-1}}-\frac{k_{N-1}}{m_{N-1}})  \\
   \dots  &  0  &   (- z\frac{c_{N-1}}{m_N}- \frac{k_{N-1}}{m_N})  &  \alpha_{N-1}  \\
\end{array}\right] 
\end{equation}  
%%%%%%%%%%%%%%%%%%
%%%%%%%%%%%%%%%%%
%%%%%%%%%%%%%%%%%
\begin{equation}                                     \label{eq:13Z222bis} 
\left[ \begin{array}{cccc}
  \dots  &   \dots   &  \dots   & \dots   \\
  \dots  &   0  &   0  &   0    \\
  \dots  &   0  &   0  &   0    \\
  \dots  & ( - z\frac{c_{N-3}}{m_{N-3}} - \frac{k_{N-3}}{m_{N-3}})  &   0  &  0  \\
  \dots  &   \alpha_{N-3}   &  (-z\frac{c_{N-2}}{m_{N-2}}- \frac{k_{N-2}}{m_{N-2}}) & 0       \\
  \dots  & (-z\frac{c_{N-2}}{m_{N-1}}-\frac{k_{N-2}}{m_{N-1}})  &   \alpha_{N-2}   & 0 \\
  \dots  &  0  &   (- z\frac{c_{N-1}}{m_{N}}- \frac{k_{N-1}}{m_{N}})  &  ( - z\frac{c_{N}}{m_{N}} - \frac{k_{N}}{m_{N}})  \\
\end{array}\right] 
\end{equation}  
%%%%%%%%%%%%%%%%%%

For instance, for $\,N=4\,$ we have
%%%%%%%%%%%%%%%%%%%%%%%%%%%%%%%%%%%%%%%%%%%%%%%%%%%
%
%%%%%%%%%%%%%%%%%
\begin{equation}                                     \label{eq:13Z122} 
Z_1 \,=\, \left[ \begin{array}{cc}
   (z^2 +z\frac{c_1}{m_1}+\frac{k_1}{m_1}) &  (-z\frac{c_1}{m_1}- \frac{k_1}{m_1})   \\
    (-z\frac{c_1}{m_2}-\frac{k_1}{m_2})  &  (z^2 + z\frac{c_1 + c_2}{m_2}+ \frac{k_1 + k_2}{m_2})   \\
     0  &   (- z\frac{c_2}{m_3}- \frac{k_2}{m_3})   \\
     0  &   0  \\
\end{array}\right] 
\end{equation}  
%%%%%%%%%%%%%%%%%%

%\vspace{.4cm}

%%%%%%%%%%%%%%%%%
\begin{equation}                                     \label{eq:13Z22244} 
Z_2\,=\, \left[ \begin{array}{ccc}
              0  &   0                                            \\
              (- z\frac{c_2}{m_2} - \frac{k_2}{m_2})  &   0         \\
    (z^2 +z\frac{c_2+c_3}{m_3}+\frac{k_2+k_3}{m_3}) &  (-z\frac{c_3}{m_3}- \frac{k_3}{m_3})     \\
    (-z\frac{c_3}{m_4}-\frac{k_3}{m_4})  &  (z^2 + z\frac{c_3 + c_4}{m_4}+ \frac{k_3 + k_4}{m_4})   \\
\end{array}\right] 
\end{equation}  
%%%%%%%%%%%%%%%%%%

%%%%%%%%%%%%%%%%
\begin{equation}                                     \label{eq:13Z2233333333} 
Z_3\,=\, \left[ \begin{array}{ccc}
              0  &   0                                            \\
              (- z\frac{c_2}{m_2} - \frac{k_2}{m_2})  &   0         \\
    (z^2 +z\frac{c_2+c_3}{m_3}+\frac{k_2+k_3}{m_3}) &  0     \\
    (-z\frac{c_3}{m_4}-\frac{k_3}{m_4})  &  (-z\frac{c_4}{m_4}- \frac{k_4}{m_4})   \\
\end{array}\right] 
\end{equation}  
%%%%%%%%%%%%%%%%%%

%%%%%%%%%%%%%%%%%%%%%%%%%%%%%%%%%%%%%%%%%%%%%%%%%%%%%%%%%%%%%%%%%%%%%%%%%%%%%%%%%%%%%%%%%%%%%%%%%%%
\begin{equation}                                     \label{eq:13ZZbis} 
\{1\}_4 =  \left[ \begin{array}{cccc}
  (z^2 +z\frac{c_1}{m_1}+\frac{k_1}{m_1}) &  (-z\frac{c_1}{m_1}- \frac{k_1}{m_1})     &          0          &          0     \\
 (-z\frac{c_1}{m_2}-\frac{k_1}{m_2})  &  \alpha_{1} & (- z\frac{c_2}{m_2} - \frac{k_2}{m_2})  &   0   \\
 0  &   (- z\frac{c_2}{m_3}- \frac{k_2}{m_3})  & \alpha_{2} &  (-z\frac{c_3}{m_3}- \frac{k_3}{m_3})     \\
  0  &   0  &  (-z\frac{c_3}{m_4}-\frac{k_3}{m_4})  &  \alpha_{3}   \\
                          \end{array}\right] 
\end{equation}
 %%%%%%%%%%%%%%%%%%%%%%%%%%%%%%%%%%%%%%%%%%%%%%%%%%%%%%%%%%%%%%%%%%%%%%%%%%%%%%%%%%%%%%%%%%%%%%%%%%%%%%
and

%%%%%%%%%%%%%%%%%%%%%%%%%%%%%%%%%%%%%%%%%%%%%%%%%%%%%%%%%%%%%%%%%%%%%%%%%%%%%%%%%%%%%%%%%%%%%%%%%%%
\begin{equation}                                     \label{eq:13ZZtris} 
\{2\}_4 =  \left[ \begin{array}{cccc}
  (z^2 +z\frac{c_1}{m_1}+\frac{k_1}{m_1}) &  (-z\frac{c_1}{m_1}- \frac{k_1}{m_1})     &          0          &          0     \\
 (-z\frac{c_1}{m_2}-\frac{k_1}{m_2})  &  \alpha_{1} & (- z\frac{c_2}{m_2} - \frac{k_2}{m_2})  &   0   \\
 0  &   - z\frac{c_2}{m_3}- \frac{k_2}{m_3}  & \alpha_{2} &  0     \\
  0  &   0  &  (-z\frac{c_3}{m_4}-\frac{k_3}{m_4})  &  (-z\frac{c_4}{m_4}-\frac{k_4}{m_4})   \\
                          \end{array}\right] 
\end{equation}
 %%%%%%%%%%%%%%%%%%%%%%%%%%%%%%%%%%%%%%%%%%%%%%%%%%%%%%%%%%%%%%%%%%%%%%%%%%%%%%%%%%%%%%%%%%%%%%%%%%%%%%

\vspace{.4cm}

%%%%%%%%%%%%%%%%%%%%%%%%%%%%%%%%%%%%%%%%%%%%%%%%%%% \framebox[2.6in]

%%%%%%%%%%%%%%%%%%%%%%%%%%%%%%%%%%%%%%%%%%%%%%%%%%% 
Now let

\begin{equation}                                     \label{eq:16.2} 
E_N \, =\,  \left[ \begin{array}{cccccc}
         (z^2 + z\frac{c_1}{m_1}+ \frac{k_1}{m_1})   &  \dots  &  .  &  .  &   0  \\
         .  &  \dots  &  .  &   .  &   0  \\
         0 &  \dots  &  .  &  .  &  .   \\
         0  &   \dots  &  .    &  (-z \frac{c_{N-2}}{m_{N-2}} - \frac{k_{N-2}}{m_{N-2}})  &  0 \\
         .  &  \dots  &  .  &   \alpha_{N-2} &    0  \\
         0  &  \dots  &   0  &   (- z\frac{c_{N-1}}{m_N} - \frac{k_{N-1}}{m_N})  &  (z\frac{c_N}{m_N} +\frac{k_N}{m_N})  \\
\end{array} \right] \; .  \end{equation}

Since 
$$z^2 I_N + z C_N + K_N$$
\begin{equation}                                     \label{eq:16.1} 
= \left[ \begin{array}{cccccc}
         (z^2 + z\frac{c_1}{m_1}  + \frac{k_1}{m_1})  &  \dots   &   0  &   0  & \dots\\
         .  &   \dots    &   .  &   .  & \dots \\
         .  &   \dots   &  .  &  .  & \dots  \\
         .  &    \dots  &  0  &  0  & \dots\\  
         .  &   \dots  & \alpha_{N-2} &   
         (- z\frac{c_{N-1}}{m_{N-1}}  - \frac{k_{N-1}}{m_{N-1}} )  & 0  \\
         0  &  .    &  0  &  ( - z\frac{c_{N-1}}{m_N}- \frac{k_{N-1}}{m_N}) & (z^2 + z\frac{c_{N-1}}{m_N}+ \frac{k_{N-1}}{m_N}) \\
\end{array} \right] 
\end{equation}

by the linearity property of determinants with respect to columns, we can express  $\, \left\| \{1\}_N \right\|\, $ within (\ref{eq:12}) as

\begin{equation}                                     \label{eq:15} 
\left\|\{1 \}_N\right\| = \left\|z^2 I_N + z C_N+ K_N\right\| + \left\|E_N\right\| = P_N (z) + \left\|E_N\right\| \;  ,
\end{equation}

Again by the linearity property of determinants we have $\, \left\|E_N\right\| + \left\|\left\{2\right\}_N\right\| =0\,$  since the last column  of the matrix
$\, E_N  +  \left\{2\right\}_N \,$ vanishes; hence, taking into account equalities  (\ref{eq:111}) through  (\ref{eq:16.2}), equality (\ref{eq:12}) yields 

\begin{equation}                                     \label{eq:17} 
P_{N+1} (z) = \left(  z^2 + z\frac{c_N}{m_{N+1}}+ \frac{k_N}{m_{N+1}} \right) P_N (z)+ z^2 \left\|E_N\right\| \; .
\end{equation}
Replacing determinant $\left\|E_N\right\|$ by its expansion in the cofactors of the $n$th column in equalities (\ref{eq:17}) and (\ref{eq:15}), respectively yields
\begin{equation}                                     \label{eq:18} 
P_{N+1} (z) = \left(  z^2 + z\frac{c_N}{m_{N+1}}+ \frac{k_N}{m_{N+1}} \right) P_N (z) +  
          z^2 (z\frac{c_N}{m_N}+\frac{k_N}{m_N}) \left\|\left\{1\right\}_{N-1}\right\|  \end{equation}

and
\begin{equation}                                     \label{eq:19} 
\left\|\left\{1\right\}_N\right\| = P_N (z) + z^2 (z\frac{c_N}{m_N}+\frac{k_N}{m_N})\left\|\left\{1\right\}_{N-1}\right\| \;, 
\end{equation}

where the matrix $\,\left\{1\right\}_{N-1}\,$ is defined by  (\ref{eq:3-4}) for $\,i=N-1$.  

Thus we have proved (\ref{eq:3-1}) through (\ref{eq:3-3}).

Lastly note that by (\ref{eq:13ZZqua})- (\ref{eq:13Z222}) 
equalities  (\ref{eq:3-4}),   (\ref{eq:wwwww})  yield   (\ref{eq:3-4nn}).    

 \vspace{.5cm}
\begin{flushright}
Q.E.D.
\end{flushright}
 \vspace{.5cm}

\begin{remark}                                \label{rem1} 
By the above inductive definition, the polynomial $\, P_N(z) \,$ can be written in the form
\vspace{.2cm}

$$P_N(z)=\Big( z^2+\frac{zc_{N-1}+k_{N-1}}{m_N} \Big)  P_{N-1}(z) + z^2\frac{zc_{N-1}+k_{N-1}}{m_{N-1}} 
 \Big(  P_{N-2}(z)+ \ldots $$
\begin{equation}                              \label{eq:3primo} 
\ldots + \frac {zc_3+k_3}{m_3} \Big(P_2 (z) + \frac {zc_2+k_2}{m_2} \big(z^2+\frac{zc_1+k_1}{m_1} \big) \ldots \Big). 
\end{equation} 
Indeed, replacing back with equalities (\ref{eq:3-1})-(\ref{eq:3-2}) in (\ref{eq:3-3}) yields (\ref{eq:3primo}).

Moreover,  the above inductive definition implies the following expression for $\, P_N(z) \,$:   
%%%
\begin{equation}                               \label{eq:3secondo} 
P_N(z)=\sum_{\rho=0}^{2N-2} \sum_{    \stackrel{2M+R=0}{0 \leq M \,;\, R \leq \rho}} 
\left( \sum_{    \stackrel{ \stackrel{ 1\leq r_1<r_2<\ldots<r_{M}<N}{{1\leq s_1<s_2< \ldots<s_{R}<N}} } {1\leq t_1<t_2<\ldots<t_{M+R}\leq N} }  %\hspace{.3cm} 
 \frac{k_{r_1} k_{r_2} \ldots k_{r_M} c_{s_1} c_{s_2} \ldots c_{s_R}} {  m_{t_1} m_{t_2} \ldots m_{t_{M+R}}} \right) \,  z^{2N-\rho}   \, .
\end{equation} 
 
\end{remark}   
%%%%%%%%%%%%%%%%%%%%%%%%%%%%%%%%%%%%%%%%%%%%%%%

For example, 
for $N=2$ we have  $\quad  P_2 (z) = \left(  z^2 + \frac{zc_1+k_1}{m_2} \right)P_1 (z) +  z^2 \frac{zc_1+k_1}{m_1}D_0 \, , $
thus
$$  P_2 (z) =  z^4 + z^3\left(\frac {c_1}{m_1} + \frac{c_1}{m_2} \right) + z^2\left(\frac {k_1}{m_1} + \frac{k_1}{m_2} \right) \, . $$
\vspace{.5cm}

For $N=3$ we have 

      $$P_3 (z) =  \left( z^2 + \frac{zc_2+k_2}{m_3} \right) P_2 (z) +
         z^2 \frac{zc_2+k_2}{m_2} \left( P_1 (z) + \frac{zc_1+k_1}{m_1} D_0 \right) \, , $$
         
thus

   $$P_3 (z) =  \left( z^2 + \frac{zc_2+k_2}{m_3} \right) \left[ z^4 + z^3\left(\frac {c_1}{m_1} + \frac{c_1}{m_2} \right) + z^2\left(\frac {k_1}{m_1} + \frac{k_1}{m_2} \right)\right] +
         z^2 \frac{zc_2+k_2}{m_2} \left( z^2 + \frac{zc_1+k_1}{m_1} \right) \, , $$

 and 

      $ P_3 (z) =  z^6 + 
                   z^5  \left(  \frac{c_1}{m_1} + \frac{c_1}{m_2} +  \frac{c_2}{m_2} + \frac{c_2}{m_3}\right)
     +  z^4  \left( \frac{k_1}{m_1} +  \frac{k_1}{m_2} + \frac{k_2}{m_2} + \frac{k_2}{m_3} +
             \frac{c_1c_2}{m_1m_2} + \frac{c_1c_2}{m_1m_3}+\frac{c_1c_2}{m_2m_3}               \right)  $
               
    $ +  z^3 \left(   \frac{k_1c_2}{m_1m_2}  + 
                      \frac{k_1c_2}{m_2m_3}  + \frac{k_2c_1}{m_1m_2}  +\frac{k_1c_2}{m_1m_3}  + \frac{k_2c_1}{m_1m_3} +  \frac{k_2c_1}{m_2m_3}  \right)         +  z^2 \left(   \frac{k_1k_2}{m_1m_2}  + \frac{k_1k_2}{m_2m_3} + \frac{k_1k_2}{m_1m_3}  \right) \, .$  
%\vspace{.5cm}
\medskip

One may verify that the same expressions can be obtained by using  (\ref{eq:3secondo}).

\section{On controllability and observability of the physical systems considered}
%Now we show that the adjoint polymomial and the characteristic polynomial have no common root.
% This result, by Theorem \ref{th:uno}, implies complete controllability and observability of the physical system.

\begin{theorem}                                                                   \label{th:unoaa}
For $N=2$ the mechanical system $\Sigma$ is completely reachable and completely observable for any choice of $\,m_1, \; m_2, \; c_1$  and $ k_1$.
\end{theorem}        
         
\underline{Proof.}$\;$  In connection with the square matrix $\,F_2\,$ of order $4$ defined in 
(\ref{eqnarray:eq11bcd}), consider the characteristic polynomial $P_2(z)$ and  the adjoint polynomial
$\,h_2\,adj(a_2)\,g_2\,$, respectively  defined by equalities (\ref{eq:4})  and (\ref{eq:matrixaN}) through (\ref{eq:hhhbxx}).

By Theorem \ref{th:tcp} and  Theorem \ref{th:adjpol}, we respectively have 
$$P_2(z)=z^4+(zc_1+k_1)( \frac{1}{m_1}+\frac{1}{m_2} )z^2 \, , \qquad h_2\,adj(a_2)\,g_2(z)=\frac{zc_1+k_1}{m_1m_2} \, .$$
  Hence  by replacing in $P_2(z)$ the unique root $z_1=-k_1/c_1$  of $h_2\,adj(a_2)\,g_2(z)$, we have
$P_2(z_1)=z_1^4=(\frac{k_1}{m_1})^4 \neq 0$.  Thus the characteristic polynomial and the adjoint polynomial have no common root and Theorem \ref{th:uno} implies that $\Sigma$ is completely reachable and completely observable;
moreover, by assertions $(E)$, $(F)$ of Section  \ref{subsection:msmch},  $\, \Sigma$ is completely controllable and completely re-construictible too.
  \begin{flushright}
Q.E.D.
\end{flushright}

\begin{theorem}                                     \label{th:adjpolnn}
Let $\,N\,$ be any integer $\,>2$.  If 
\begin{equation} \label{eq:hyp-prop} c_1/k_1=c_2/k_2=\ldots =c_{N-1}/k_{N-1} \; , \end{equation}
 then 
 
 $(A)\;$  the mechanical system $\Sigma$ is completely reachable and completely observable for any choice of 
$\,m_1, \, m_2, \ldots \, m_N$.
  \end{theorem} 

\underline{Proof.} $\;$  By Theorem \ref{th:adjpol} the roots of the adjoint polynomial (\ref{eq:hhhbxx}) are 
$z_i=-k_i/c_i ,  \;  i=1 , \dots,  \, N-1. \,$  Let 

\begin{equation}                                             \label{eq:xxyyzzuu}
	b_i  =	b_i(z)   = z c_i + k_i \, \qquad \beta_{ij}=-k_j\frac{c_i}{c_j}+k_i \,  ,
\end{equation}
hence $\, b_i=0\,$ if and only if $\,z=z_i$.  By hypothesis (\ref{eq:hyp-prop}) we have 
\begin{equation}                                             \label{eq:xxyyzzuu22222222}
\beta_{ij}=0 \,  , \qquad  i,\,j=1,\,2,\ldots,\, N-1 \, .
\end{equation}

Hence the inductive definition (\ref{eq:3-1})-(\ref{eq:3-3}) of $\,P_N(z)\,$ yields
\begin{equation} \label{eq:3-26} 
P_N(z_j)=z_j^2P_{N-1}(z_j)=z_j^4P_{N-2}(z_j)=\ldots=(k_j/c_j)^{2N}\neq 0\, .
\end{equation}

Thus any root $z_j$ of the adjoint polynomial cannot be a root for the characteristic polynomial too.

Now, Theorem \ref{th:uno} implies that $\Sigma$ is completely reachable and completely observable;
and by assertions $(E)$, $(F)$ of Section  \ref{subsection:msmch},  $\Sigma$ is completely controllable and completely re-construictible too.
\begin{flushright}
Q.E.D.
\end{flushright}

In particular, by the above theorem, for $N=3$ the mechanical system $\Sigma$ is completely reachable and completely observable for any choice of 
$\,m_1$, $\, m_2$, $\, m_3$, $\, c_1$, $\, k_1$, $\,c_2\,$ and $\, k_2\,$ provided that $\,c_1/k_1=c_2/k_2$ .

The next two theorems respectively show that when  the sequences $\,\left\{c_i\right\}\,$ and $\,\left\{k_i\right\}\,$ are not proportional
proposition $(A)$ can be either true or not true.
Hence, in particular, condition (\ref{eq:hyp-prop}) is  sufficient  but not necessary for the validity of  $(A)$.

   \begin{theorem}                                     \label{th:notsuff}
Proposition $(A)$ is not true for some mechanical system $\Sigma$ such that (\ref{eq:hyp-prop}) does not hold. 
  \end{theorem} 

\underline{Proof.}$\;$  Let  $N=3$.  We  have 

      $ P_3 (z) =  ( z^2 + \frac{b_2}{m_3} ) P_2 (z) +
         z^2 \frac{b_2}{m_2} ( P_1 (z) + \frac{b_1}{m_1} ) \, , $

 {\it i.e.},         
         
   \begin{equation}                                     \label{eq:ptrebbblll}
 P_3 (z) =  \left( z^2 + \frac{b_2}{m_3} \right) \left(  z^4 + b_1 \frac{1}{m_2}z^2 +  
 z^2 \frac{b_1}{m_1}\right) + z^2 \frac{b_2}{m_2} \left( z^2 + \frac{b_1}{m_1} \right) \, , 
    \end{equation}

Thus, if $\,z_1=0\,$ is root of the adjoint polynomial, {\it i.e.}

if $\,b_1=0\,$,  then   $\,P_3 (z_1) =  z_1 ^6  + \beta_{12}H z_1^4 \, ,$  with     $\,H=\frac{1}{m_2} + \frac{1}{m_3}$.

Hence $\,P_3(z_1) =  0 \,$   if and only if   $\,z_1^2 + \beta_{21}H =0\,$, {\it i.e.},   

$$\frac{k_1^2}{c_1^2}+(k_2 - \frac{k_1}{c_1} c_2)H=0$$
 
This relation is linear in $k_2$;  hence, if $H$, $k_1$, $c_1$, $c_2$ are given, then one determines the unique $k_2$ such that the last equality holds. 
With regard to these values of the material parameters it follows that $\,z_1$ is a root for both the characteristic polynomial and adjoint polynomial.
\begin{flushright}
Q.E.D.
\end{flushright}

   \begin{theorem}                                     \label{th:notnecessary}
Proposition $(A)$ is  true for some mechanical system $\Sigma$ such that (\ref{eq:hyp-prop}) does not hold. 
  \end{theorem} 

\underline{Proof.}$\;$  Let  $N=3$.  By (\ref{eq:ptrebbblll}) we have 
   \begin{equation}                                     \label{eq:ptrebbbxx}
 P_3 (z_1) =  \left( z_1^2 + \frac{b_2(z_1)}{m_3} \right) z_1^4 + z_1^2 \frac{b_2(z_1)}{m_2}  z_1^2 =
  \frac{k_1^4}{c_1^4} \Big[\frac{k_1^2}{c_1^2}+(k_2-\frac{k_1}{c_1}c_2)H_{23}\Big]\, , 
    \end{equation}  
   \begin{equation}                                     \label{eq:ptrebbbccc}
 P_3 (z_2) =  z_2^4 \left( z_2^2 + \frac{b_1(z_2)}{m_3} \right) z_1^4 + z_1^2 \frac{b_2(z_1)}{m_2}  z_1^2 =
  \frac{k_2^4}{c_2^4} \Big[\frac{k_2^2}{c_2^2}+(k_1-\frac{k_2}{c_2}c_1)H_{12} \Big]\, , 
    \end{equation}  
    where
 \begin{equation}                                     \label{eq:ptrebbbccc2}
H_{ij}=\frac{1}{m_i} + \frac{1}{m_j} \,  
    \end{equation}  
    and
        \begin{equation}                                     \label{eq:ptrebbbccc3}
z_{i}=-\frac{k_1}{c_i}    
    \end{equation}  
are the roots of the adjoint polynomial.   
Hence
   \begin{equation}                                     \label{eq:ptrebbbyy}
 P_3 (z_1) =0  \; \Leftrightarrow \;  \frac{k_1^2}{c_1^2}+(k_2-\frac{k_1}{c_1}c_2)H_{23}= 0  \,  
    \end{equation}  
    and
   \begin{equation}                                     \label{eq:ptrebbbzzz}
 P_3 (z_2) =0  \; \Leftrightarrow \;  \frac{k_2^2}{c_2^2}+(k_1-\frac{k_2}{c_2}c_1)H_{12}=0  \, .
    \end{equation}
Thus, given  $m_1,\,m_2,\,m_3, \,c_1$ and $c_2$ one can  choose $(k_1,\,k_2)$ such that both  
 $P_3(z_1)\neq0$  and  $P_3(z_2)\neq0$.  Hence the adjoint polynomial and the characteristic polynomial have no common root.
 In connection with such a choice for the mechanical system $\Sigma$ proposition $(A)$ is true.
\begin{flushright}
Q.E.D.
\end{flushright}
%\newpage

%%%%%%%%%%%%%%%%%%%%%%%%%%%%%%%%%%%%%%%%%%%%%%%%%%%%%%%%%%%%%%%%%%%%%%%%%%%%%%
 \section{On applications of the above results}
 %%%%%%%%%%%%%%%%%%%%%%%%%%%%%%%%%%%%%%%%%%%%%%%%%%%%%%%%%%%%%%%%%%%%%%%%%%%%%%

The system of equations $\, \Sigma_N \,$  studied here is a mathematical model suitable to describe various rectilinear physical systems as chains of three-dimensional bodies undergoing translational motions.
  
By the results of Theorems  5.1-5.2 above, people involved in the design of such physical systems with $\, N > 2\,$ can choose the material constants in order to render the system completely controllable, observable,  reachable and re-construictible; instead for $\,N=2\,$ any such a system has such properties for any choice of the material constants.  
 
 The key advantage of the theorems proved here for engineering applications is that, when control is a property useful for the system under design, one can design the system by choosing the material parameters in such a way as to have controllability. 
  
A further theoretical task will be  to construct or to find the controlling input functions for the system.  The present paper clarifies when such a problem has a solution.

An example of  a physical system, which can be modelled by $\,\Sigma_N\,$, is given by a chain of   $\,N\,$  pistons consecutively connected by springs and constrained  inside a cylindrical cavity containing a fluid.   
Assuming that an external input  force  $\,u(\tau)\,$ can act normally on the first piston, and taking as output the position of the latter piston, the system will be controllable and observable by the history of the latter provided the elastic springs be chosen proportional to the constants of viscous damping between consecutive pistons.  Hence, (i) the system can reach any prescribed state by a suitable input force   $\,u(\tau)\,$ ;  and  (ii)  every state of the system (that is, position and velocity of each piston) can be determined by observing the excursions of the latter piston.

In particular, system $\, \Sigma_2 \,$    can also be used  to model the behaviour of  car-wheel suspensions. In more detail, when a car travels along a bumpy road the wheel tyre copies roughness of the road surface.
The wheel is thus driven up or down in the vertical direction along the \(z\)  axis.
In this case, the rectilinear system under consideration consists of four bodies: the spring, the shock absorber, the wheel and the quarter of the car body.
The system excitation by the road surface and the gravitational attraction of the wheel and the quarter-car body form the system surroundings.  
The model is excited by the source of the velocity  $\, \dot z_o \,$  resulting from the tyre copying the road bumps.

% ??**

%\setcounter{figure}{3}
%\begin{sidewaysfigure}   
%\begin{figure}[h]   % h==HERE      t=top of page
%\centerline{\epsfysize=8cm \epsfbox{quartercar.eps} \hskip 5cm  \epsfysize=8cm \epsfbox{quartercar.eps}}
%\epsfbox{quartercar.eps}
%\caption{Quarter car model}
%\label{landfig}
%\end{figure}
%\end{sidewaysfigure}

\begin{figure}
\centerline{\epsfysize=8cm\epsfbox{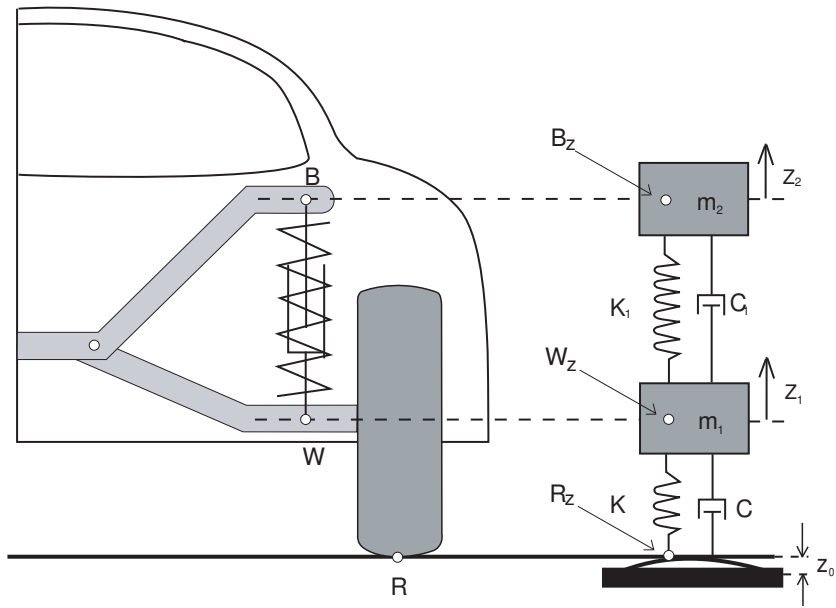}}
\caption{Quarter car model}
\label{landfig:tre}
\end{figure}

Figure \ref{landfig:tre} on the left  shows a  detail of the suspension of one of its wheels,   and   on the right  shows a scheme of it.  The point  $R_z$  represents the z-position of the contact point between the car tyre and the road surface R  when the car moves in the  x-direction.  
The points  $W_z$  and  $B_z$  represent the z-motion of the wheel axis $W$ and the body-suspension interaction point $B$, respectively.
The constants $k$  and $c$  are the compliance and damping of the wheel tyre,  $\,m_1\,$ is the mass of the complete wheel and $\,m_2\,$ is the quarter of the mass of the car body.  The model is excited by the source of the velocity  $\, \dot z_o\,$ resulting from the tyre copying the road bumps.
Then by applying Newton's second law we obtain the differential equations of  this physical model :
%%%%%%%%%%%%%%%%%%%%%%%%%%%%%%%%%%%
\begin{equation}  														         						 
m_1 {{d^2}\over{dt^2}}z_1 = k_1 (z_{2}-z_1) + c_1 (\dot{z}_{2}-\dot{z}_1) + f^{ext}_1    \; ,   \label{eq:eqm1appl}  
\end{equation}     %%%%%%%%%%%%%%%%%%%%%%%%%%%%%%%%%%%%%%%%%%%%%%%%%%%%%%%%%%%%%%%%%
\begin{equation}   \label{eq:eqmiappl}  
m_2 {{d^2}\over{dt^2}}z_2 = k_{1} (z_{1}-z_2) + c_{1}(\dot{z}_{1}-\dot{z}_2) ,  
   \end{equation}
%%%%%%%%%%%%%%%%%%%%%%%%%%%%%%%%%%%
with
%%%%%%%%%%%%%%%%%%%%%%%%%%%%%%%%%%%
\begin{equation}  														         						 
f^{ext}_1 =  k(z_{o}-z_1) + c(\dot{z}_{o}-\dot{z}_1)   \; ,   \label{eq:eqm1applfext}  
\end{equation}     %%%%%%%%%%%%%%%%%%%%%%%%%%%%%%%%%%%%%%%%%%%%%%%%%%%%%%%%%%%%%%%%%

We see that the system considered is a particular  $\, \Sigma_2\,$  system (see equations (\ref{eq:eqm1})-(\ref{eq:eqmn})).   Hence by Theorem  5.1 it is observable and re-construictible for arbitrary choices of all the material parameters. 

If  one wants to know how e.g. the wheel center  \(W\)  behaves in response to the external force road input, without a priori knowing the latter,
he may observe the output past excursions  \(\,y(s):=z_2(s)\,\)  of the body-suspension interaction point: they determine the actual state of  the system, thus the actual position of  W  too. 
Thus he can go back to the external force road input which generates the vertical motion of the suspension.

%%%%%%%%%%%%%%%%%%%%%%%%%%%%%%%%%%%%%%%%%%%%%%%%%%%%%%%%%%%%%%%%%%%%%%%%%%%%%%%%%%%%%%%%%
%%%%%%%%%%%%%%%%%%%%%%%%%%%%%%%%%%%%%%%%%%%%%%%%%%%%%%%%%%%%%%%%%%%%%%%%%%%%%%%%%%%%%%%%%
 
%%%%%%%%%%%%%%%%%%%%%%%%%%%%%%%%%%%%%%%%%%%%%%%%%%%%%%%%%%%%%%%%%%%

\newpage
%%%%%%%%%%%%%%%%%%%%%%%%%%%%%%%%%%%%%%%%%%%%%%%%%%%%%%%%%%%%%%%%%%%%%%%%%%%%%%%%%%%%%%%%%%%%%%%%%%%%%%%%%%%%%%%%%%%%%%
%\setcounter{figure}{1}
%\begin{sidewaysfigure}
%%%%%%%%%%%%%%%%%%%%%%%%%%%%%%%%%%%%%%%%%%%%
%\setcounter{figure}{2}
%\begin{sidewaysfigure}
%%%%%%%%%%%%%%%%%%%%%%%%%%%%%%%%%%%%%%%%%%%%%%
%\setcounter{figure}{3}
%\begin{sidewaysfigure}
%\end{sidewaysfigure}

\end{document}